\documentclass[aps, prd, superscriptaddress, nofootinbib, preprintnumbers, showpacs]{revtex4}
\usepackage{graphicx}
\usepackage{amsmath}
\def\fsl#1{\setbox0=\hbox{$#1$}                 
   \dimen0=\wd0                                 
   \setbox1=\hbox{/} \dimen1=\wd1               
   \ifdim\dimen0>\dimen1                        
      \rlap{\hbox to \dimen0{\hfil/\hfil}}      
      #1                                        
   \else                                        
      \rlap{\hbox to \dimen1{\hfil$#1$\hfil}}   
      /                                         
   \fi}                                         %
\newcommand{\diag}{\mbox{diag}}

\newcommand{\VEV}[1]{\langle #1 \rangle}

\preprint{UWO-TH-05/04}
\begin{document}
\title{Gluonic phase in neutral two-flavor dense QCD}

\author{E.V. Gorbar}
  \email{egorbar@uwo.ca}
  \altaffiliation[On leave from ]{
       Bogolyubov Institute for Theoretical Physics,
       03143, Kiev, Ukraine}
\author{Michio Hashimoto}
  \email{mhashimo@uwo.ca}
\author{V.A. Miransky}
  \email{vmiransk@uwo.ca}
   \altaffiliation[On leave from ]{
       Bogolyubov Institute for Theoretical Physics,
       03143, Kiev, Ukraine}
\affiliation{
Department of Applied Mathematics, University of Western
Ontario, London, Ontario N6A 5B7, Canada
}

\date{\today}

\begin{abstract}
In the Ginzburg-Landau approach, we describe a new
phase in neutral two-flavor quark matter in which
gluonic degrees of freedom play a crucial role. We call
it a gluonic phase. In this phase gluonic dynamics cure
a chromomagnetic instability in the 2SC solution
and lead to spontaneous
breakdown of the color gauge symmetry,
the electromagnetic $U(1)$, and the rotational $SO(3)$. 
In other words, the
gluonic phase 
describes an
anisotropic medium in which the color and electric
superconductivities coexist. Because most of the initial symmetries
in this system are spontaneously broken, its dynamics is very rich.
\end{abstract}


\maketitle

\section{Introduction}

It is natural to expect that
cold quark matter 
may exist in the interior of compact stars. This fact motivated 
intensive studies of this system 
over the past few years (for a review, see Ref. \cite{RW}). While
these studies firmly established that
the cold and dense quark matter is a color superconductor,
they also revealed a remarkably rich phase structure in this system,
consisting of many different phases. The question which 
phase is picked up by nature is still open.

In this Letter we will describe a new phase in neutral 
and $\beta$-equilibrated two-flavor
quark matter. We call it a gluonic phase.
The name reflects a crucial role of 
gluonic degrees of freedom in the structure of its ground state.
More precisely, besides the usual color superconducting 
condensate of quarks, there exist (vector) condensates of gluons
in this phase.
These vector condensates cure a chromomagnetic instability in the
two-flavor superconducting (2SC) solution \cite{HS} and
lead to a dramatic rearrangement of the 2SC
ground state, most
notably, to spontaneous breakdown of 
$SU(2)_c \times \tilde{U}(1)_{em}\times SO(3)_{\rm rot}$ symmetry 
down to $SO(2)_{\rm rot}$. Here $SU(2)_c$ and $\tilde{U}(1)_{em}$ are
the color and electromagnetic gauge symmetries 
in the 2SC medium, and $SO(3)_{\rm rot}$ 
is the rotational group (recall that in the 2SC solution the
color $SU(3)_c$ is broken down to $SU(2)_c$).
In other words, the gluonic phase describes an
anisotropic medium in which the color and electric 
superconductivities coexist. 
As we will discuss below, there may exist a class
of solutions with vector condensates of gluons. The solution
described in this Letter is
similar to that found in the gauged $\sigma$-model with a
chemical potential for hypercharge \cite{GMS}, although physics in
the present system is much richer. In particular, as will be shown
in Sec. \ref{5}, exotic hadronic states play an important role
in its dynamics.

The framework of the present study is the Ginzburg-Landau (GL)
approach. The basic point in our analysis is the 
inclusion of light gluonic degrees of
freedom in the GL effective action. This leads us
to revealing the gluonic phase in neutral two-flavor quark matter. 

\section{Effective potential with vector condensates}

We study dense two-flavor quark matter in $\beta$-equilibrium.
For our purpose, it is convenient to use a phenomenological 
Nambu-Jona-Lasinio (NJL) model, more precisely, a gauged NJL model
including gluons.
Although usually the NJL model is regarded as a low-energy 
effective theory in which massive gluons are integrated out,
we introduce gluonic degrees of freedom because 
the gluons of the unbroken $SU(2)_c$ subgroup of the color
$SU(3)_c$
are left as massless, and, under certain conditions considered below,
some other gluons can be also very light. As discussed in Ref.~\cite{RSS},
the confinement scale $\Lambda'_{\rm QCD}$ in the two-flavor color 
2SC phase is estimated as ${\cal O}(\mbox{10 MeV})$
or smaller.
Thus it is not peculiar to consider gluonic degrees of freedom 
at energies less than the quark gap $\Delta$.

We neglect the current quark masses and, as usual in studying the
2SC phase, assume that the color superconducting condensate
does not break parity.
The Lagrangian density is then given by
\begin{equation}
  {\cal L} = \bar{q}(i\fsl{D}+\hat{\mu}\gamma^0)q
  +G_\Delta \bigg[\,(\bar{q}^C i\varepsilon\epsilon^a\gamma_5 q)
                   (\bar{q} i\varepsilon\epsilon^a\gamma_5 q^C)\,\bigg]
  + {\cal L}_g,
 \label{Lag}
\end{equation}
where
\begin{equation}
  {\cal L}_g = -\frac{1}{4}F_{\mu\nu}^{\alpha} F^{\alpha\,\mu\nu}
\end{equation}
and
\begin{equation}
  D_\mu \equiv \partial_\mu - ig A_\mu^{\alpha} T^{\alpha}, \quad
  F_{\mu\nu}^{\alpha} 
  \equiv \partial_\mu A_\nu^{\alpha} - \partial_\nu A_\mu^{\alpha} +
  g f^{\alpha\beta\gamma} A_\mu^{\beta} A_\nu^{\gamma}.
\end{equation}
Here $A_\mu^{\alpha}$ are gluon fields, 
$T^{\alpha}$ are the $SU(3)$ matrices in 
the
fundamental representation, and 
$\varepsilon^{ij}$ and $\epsilon^{acd}$ are
the antisymmetric tensors in the flavor and color spaces, respectively.
In $\beta$-equilibrium, the
elements of the diagonal chemical potential matrix $\hat \mu$ 
for up ($u$) and down ($d$) quarks are
\begin{subequations}
\label{beta-eq}
\begin{align}
  \mu_{ur} &= \mu_{ug} = \tilde{\mu} - \delta \mu, 
  &\mu_{dr}&= \mu_{dg} = \tilde{\mu} + \delta \mu, \\
  \mu_{ub} &= \tilde{\mu} - \delta \mu - \mu_8, 
  &\mu_{db}&= \tilde{\mu} + \delta \mu - \mu_8 ,
\end{align}
\end{subequations}
with
\begin{equation}
  \tilde{\mu} \equiv \mu - \frac{\delta\mu}{3} + \frac{\mu_8}{3}, \qquad
  \delta \mu \equiv \frac{\mu_e}{2}.
\label{mu}
\end{equation}
Here the subscripts $r$, $g$, and $b$ correspond to red, green and 
blue quark colors,
$\mu$ is the quark chemical potential 
(the baryon chemical potential $\mu_B$ is given by $\mu_B \equiv 3\mu$),
$\mu_e$ is the chemical potential for the electric charge, and
$\mu_8$ is the color chemical potential.
The latter is simply connected with the vacuum expectation value
(VEV) of the time component of the 8-th gluon \cite{GR}:
\begin{equation}
  g\VEV{A_0^{8}} = \frac{2}{\sqrt{3}}\,\mu_8 .
\end{equation}

By using the auxiliary field
$\Delta^a \sim i\bar{q}^C\varepsilon \epsilon^a \gamma_5 q$,
the Lagrangian density (\ref{Lag}) can be rewritten as
\begin{equation}
  {\cal L} = \bar{q}(i\fsl{D}+\hat \mu \gamma^0)q
 -\frac{1}{2}\Delta^a[i\bar{q}\varepsilon\epsilon^a \gamma_5 q^C]
 -\frac{1}{2}[i\bar{q}^C\varepsilon\epsilon^a\gamma_5 q]\Delta^{*a}
 - \frac{|\Delta^a|^2}{4G_\Delta}
 +{\cal L}_g .
 \label{Lag_aux}
\end{equation}

We now introduce the Nambu-Gor'kov spinor, 
\begin{equation}
  \Psi = \left(\begin{array}{@{}c@{}} q \\ q^C \end{array}\right) .
\end{equation}
The inverse propagator $S_g^{-1}$ of $\Psi$ including gluons is 
written as
\begin{equation}
  S_g^{-1} = \left(
  \begin{array}{cc}
  [G_{0,g}^+]^{-1} & \Delta^- \\ \Delta^+ &  [G_{0,g}^-]^{-1} 
  \end{array}
  \right) ,
  \label{S-inv}
\end{equation}
with
\begin{equation}
  [G_{0,g}^+]^{-1} \equiv
  (p^0+\tilde{\mu}-\delta\mu\tau^3-\mu_8{\bf 1}_b)\gamma^0
  -\vec \gamma \cdot \vec p + g \fsl{A}^{\alpha}T^{\alpha},
\end{equation}
\begin{equation}
  [G_{0,g}^-]^{-1} \equiv
  (p^0-\tilde{\mu}+\delta\mu\tau^3+\mu_8{\bf 1}_b)\gamma^0
  -\vec \gamma \cdot \vec p - g \fsl{A}^{\alpha}T^{\alpha}{}^T
\end{equation}
and
\begin{equation}
  \Delta^- \equiv -i\epsilon^b\varepsilon\gamma_5 \Delta, \qquad
  \Delta^+ \equiv -i\epsilon^b\varepsilon\gamma_5 \Delta^*.
\end{equation}
Here the diquark condensate is chosen along the blue color direction, 
the constant fields $A^{\alpha}_{\mu}$ represent possible vector
gluonic condensates in the model, and
we introduced matrices $\tau^3 \equiv \mbox{diag}(1,-1)$ and
${\bf 1}_b \equiv \mbox{diag}(0,0,1)$ acting in the flavor and color 
spaces, respectively.
The effective potential in this model includes both gluons and 
the scalar field $\Delta$. It is:  
\begin{equation}
  V_{\rm eff} =
   \frac{|\Delta|^2}{4G_{\Delta}}
  +\frac{g^2}{4}f^{\alpha\beta\gamma}f^{\alpha\delta\sigma}
  A_\mu^{\beta} A_\nu^{\gamma} A^{\delta\,\mu} A^{\sigma\,\nu}
  -\frac{1}{2}\int\frac{d^4 p}{i(2\pi)^4}\ln\det S_g^{-1} .
  \label{V}
\end{equation}

We will utilize the hard dense loop approximation, in which only
the dominant one-loop quark contribution is taken into account,
while the contribution of gluon loops is neglected. On the other hand,
we keep the tree contribution of gluons in the 
effective potential (\ref{V}). This is because we want to compare 
this contribution with that of hard dense loops in order to check the
consistency of the hard dense loop approximation.

\section{Dynamics of gluons in the Ginzburg-Landau approach}

Let us consider dynamics of light gluonic degrees of freedom
in this model.
The two point function of gluons can be calculated from 
Lagrangian density (\ref{Lag_aux}).
In Refs.~\cite{HS,R},
the Debye and Meissner screening masses of the gluons in 
the 2SC phase were explicitly calculated.
For the gluons of the unbroken $SU(2)_c$, i.e., 
$A^{(1)}$, $A^{(2)}$, and $A^{(3)}$, both Debye and Meissner 
masses  
vanish in the region $\delta\mu < \Delta$
(henceforth, for clarity, we will put color indices of gluon fields
in parentheses).
For the gluons $A^{(4)-(7)}$, the Meissner mass is approximately
\begin{equation}
 m_{M,4}^2 = \frac{g^2\tilde{\mu}^2}{6\pi^2}
 \left(\,1-\frac{2\delta\mu^2}{\Delta^2}\right) , \quad \delta\mu < \Delta.
\label{m_M4}
\end{equation}
Thus, near the critical point $\delta\mu=\Delta/\sqrt{2}$,
the Meissner mass for $A^{(4)-(7)}$ is very small. As $\delta\mu$
exceeds the value $\Delta/\sqrt{2}$,  $m_{M,4}^2$ becomes negative,
thus signalizing a chromomagnetic instability in the 2SC solution 
\cite{HS}.
On the other hand, around the critical point $\delta\mu=\Delta/\sqrt{2}$,
the $SU(2)_c$ singlet gluon $A^{(8)}$ is heavy.
Actually, it has
nonvanishing Debye and Meissner screening masses in the whole region 
$\delta\mu < \Delta$.
This fact allows us to pick up the gluons $A^{(1)-(7)}$ as relevant light
degrees of freedom in the low energy effective theory around 
the critical point $\delta\mu = \Delta/\sqrt{2}$. Our goal is
to describe the dynamics near this critical point in the
GL approach.

The quark gap $\Delta \ne 0$ breaks the QCD symmetry $SU(3)_c$ 
down to $SU(2)_c$. With respect to $SU(2)_c$, 
the adjoint representation of $SU(3)_c$ is decomposed as
\begin{equation}
  {\bf 8}={\bf 3}\oplus{\bf 2}\oplus\bar{{\bf 2}}\oplus{\bf 1},
\end{equation}
i.e., 
\begin{equation}
 \{A_\mu^{(\alpha)}\} = (A_\mu^{(1)},A_\mu^{(2)},A_\mu^{(3)})
 \oplus K_\mu \oplus K_\mu^\dagger
 \oplus A_\mu^{(8)} , \qquad (\alpha=1,2,\cdots,8) .
\end{equation}
Here we defined the complex doublet of the ``matter'' field 
describing color vector ``kaons":
\begin{equation}
  K_\mu \equiv \frac{1}{\sqrt{2}}
  \left(
  \begin{array}{c}
  A_\mu^{(4)}-iA_\mu^{(5)} \\[2mm] A_\mu^{(6)}-iA_\mu^{(7)} 
  \end{array}
  \right) .
\end{equation}
We also define
\begin{equation}
  f_{\mu\nu}^{(l)} \equiv \partial_\mu A_\nu^{(l)} - 
  \partial_\nu A_\mu^{(l)}
  +g \epsilon^{lmn} A_\mu^{(m)} 
  A_\nu^{(n)}, \qquad 
  (l, m, n=1,2,3)
\end{equation}
and
\begin{equation}
  {\cal D}_\mu \equiv \partial_\mu - ig A_\mu^{(l)} 
  \frac{\sigma^l}{2}.
\end{equation}
Then the building blocks of the effective action are the following six,
\begin{equation}
  K_0, \quad K_j, \quad {\cal D}_0, \quad {\cal D}_j,  \quad 
  f_{0j}, \quad f_{jk}.
\label{blocks}
\end{equation}
The effective action should be of course invariant under all
initial symmetries, in particular, under the color $SU(2)_c$ and
the rotational $SO(3)_{\rm rot}$.

Because of the chromomagnetic instability at 
$\delta\mu > \Delta/\sqrt{2}$, it is natural to study
a spontaneous breakdown of the $SU(2)_c$ via the formation of 
a vector condensate $\VEV{K_\mu} \ne 0$. Because the instability is
chromomagnetic,
we assume that a spatial component of $K_\mu$ has a VEV.
By using the rotational symmetry $SO(3)_{\rm rot}$, one can take
$\VEV{K_{3}} \ne 0$. 
And because of the $SU(2)_c$ symmetry,
without loss of generality, we can choose $\VEV{A^{(6)}_{3}} \ne 0$.
This VEV breaks both the $SU(2)_c$ and the $SO(3)_{\rm rot}$.

The following remarks are in order. The complex doublet $K_{3}$ plays
here the role of a Higgs field responsible for spontaneous
breakdown of the $SU(2)_c$. The situation is similar to that
taking place in the electroweak theory. The essential difference
however is that now the Higgs field is a spatial component of the
vector field leading also to spontaneous breakdown of the
rotational symmetry. In this paper, the unitary gauge will be
used in which
$K_{3}^T = \frac{1}{\sqrt{2}}(0, \VEV{A^{(6)}_{3}} + a^{(6)}_{3})$   
where the real field $a^{(6)}_{3}$ describes quantum fluctuations. The 
important point is that in the unitary gauge all auxiliary 
(gauge dependent) degrees of freedom are removed. {\sl Therefore in this
gauge the vacuum expectation values of vector fields are
well-defined physical quantities.}

With a broken $SU(2)_c$, the $SU(2)_c$ gluons could have 
VEVs. A similar situation takes place in the gauged $\sigma$-model
with a chemical potential for hypercharge, where the gauge
$SU(2)_L$ is broken \cite{GMS}. Motivating by that model,
we assume
\begin{equation}
  \VEV{A^{(1)}_{3}}, \;\VEV{A^{(3)}_{0}} \ne 0,
\end{equation}
and use the following notation,
\begin{equation}
  B \equiv g\VEV{A^{(6)}_{3}}, \quad
  C \equiv g\VEV{A^{(1)}_{3}}, \quad
  D \equiv g\VEV{A^{(3)}_{0}} 
  \label{def-BCD}
\end{equation}
(note that $g\VEV{A^{(3)}_{0}}$ can be considered as a chemical
potential $\mu_3$ related to the third component of
the color isospin). As will be shown in Sec. \ref{4}, such a
solution with nonzero $B, C$ and $D$ vector condensates
exists in this model indeed.

We now describe a general symmetry breaking structure in this model.
In the presence of $\mu_e$, the chiral symmetry is explicitly broken.
Then the initial symmetry is
\begin{equation}
 [SU(3)_c]_{\rm local} \times  
 [U(1)_{em} \times
 U(1)_{\tau^3_L} \times U(1)_{\tau^3_R}]_{\rm global} \times SO(3)_{\rm rot},
\end{equation}
where $U(1)_{\tau^3_{L,R}}$ are $U(1)$-part of the chiral symmetry
$SU(2)_{L,R}$ (note that there is no photon field in the model).
The baryon charge is incorporated in the subgroup,
\begin{equation}
  {\cal B} = \frac{1}{3} {\bf 1}_f \otimes {\bf 1}_c = 2(Q-I_3),
\end{equation}
where $Q=\diag(2/3,-1/3)$ and $I_3=\diag(1/2,-1/2)$ acting on 
the flavor space.
The quark gap $\Delta$ breaks both 
$SU(3)_c$ (down to $SU(2)_c$) and $U(1)_{em}$ but
a linear combination of $T^8$
and $Q$ remains
unbroken. The new electric charge of the unbroken 
$\tilde{U}(1)_{em}$ is
\begin{equation}
  \tilde{Q} = Q-\frac{1}{\sqrt{3}}T^8.
\end{equation}
The baryon charge is also changed to
\begin{equation}
  \tilde{\cal B} = 2(\tilde{Q} - I_3) .
\end{equation}
The VEV $\VEV{A^{(6)}_{3}}$ breaks $SU(2)_c$, but a linear combination
of the generator $T^3$ from the
$SU(2)_c$ and $\tilde{Q}$,
\begin{equation}
  \tilde{\tilde{Q}} = \tilde{Q} - T^3 = Q-\frac{1}{\sqrt{3}}T^8 - T^3,
\end{equation}
determines the unbroken $\tilde{\tilde{U}}(1)_{em}$ (the new baryon
charge is
${\tilde{\tilde{\cal B}}} = 2(\tilde{\tilde{Q}} - I_3)$).
However, because $T^1$ does not commute with $T^3$,
the VEV $\VEV{A^{(1)}_{3}}$ breaks $\tilde{\tilde{U}}_{em}(1)$. The
baryon charge is also broken.

After all, we have:
\begin{eqnarray}
\lefteqn{\hspace*{-2cm}
 [SU(3)_c]_{\rm local} \times [U(1)_{em} \times 
 U(1)_{\tau^3_L} \times U(1)_{\tau^3_R}]_{\rm global} \times SO(3)_{\rm rot} 
} \nonumber \\
&& \stackrel{\Delta}{\longrightarrow}
 [SU(2)_c]_{\rm local} \times [\tilde{U}(1)_{em} \times 
 U(1)_{\tau^3_L} \times U(1)_{\tau^3_R}]_{\rm global} \times SO(3)_{\rm rot}\\
&& \stackrel{\VEV{A^{(6)}_{3}}}{\longrightarrow}
 [\tilde{\tilde{U}}(1)_{em} \times 
 U(1)_{\tau^3_L} \times U(1)_{\tau^3_R}]_{\rm global} \times SO(2)_{\rm rot}\\
&& \stackrel{\VEV{A^{(1)}_{3}}}{\longrightarrow}
 [U(1)_{\tau^3_L} \times U(1)_{\tau^3_R}]_{\rm global} \times SO(2)_{\rm rot} .
\end{eqnarray}
Thus, this system describes an anisotropic medium in which both the
color and electric superconductivities coexist.

Let us apply the GL approach to this system near the critical point
$\delta\mu \simeq \Delta/\sqrt{2}$. The $SU(2)_c$ and
$SO(3)_{\rm rot}$ 
symmetries dictate that the
general GL effective potential, made from building blocks
(\ref{blocks}) and
including operators up to the mass dimension four, 
is
\begin{equation}
  V_{\rm eff} = V_\Delta + \frac{1}{2}M_B^2 B^2
  + T D B^2 + \frac{1}{2}\lambda_{BC} B^2 C^2
  + \frac{1}{2}\lambda_{BD} B^2 D^2
  + \frac{1}{2}\lambda_{CD} C^2 D^2 
  + \frac{1}{4}\lambda_B B^4,
  \label{LG-pot}
\end{equation}
where $V_\Delta$ is the 2SC part of the effective potential.
Here, while the coefficients $\lambda_B$,
$\lambda_{BC}$, $\lambda_{BD}$, and $\lambda_{CD}$ are dimensionless,
the dimension (in mass units) of the coefficient $T$ in the triple vertex  
is one.
Expanding the potential (\ref{V}) with respect to $B$, $C$, 
and $D$, we can determine the potential (\ref{LG-pot}).
Amazingly, the structure of the gluonic 
part of effective potential (\ref{LG-pot}) 
is quite
similar to that of the potential in the gauged $\sigma$-model with the 
hypercharge
chemical potential \cite{GMS}, although the present system is much
richer.

Before realizing explicit calculations,
we clarify the behavior of the effective potential
(\ref{LG-pot}) near the critical point.
The stationary point of the effective potential (\ref{LG-pot})
is given by equations
\begin{eqnarray}
 \frac{\partial V_{\rm eff}}{\partial B} &=&
 B \left[\,M_B^2+\lambda_B B^2+2T D + \lambda_{BC}C^2 +
           \lambda_{BD} D^2\,\right] = 0,
 \label{gap-eq-B} \\
 \frac{\partial V_{\rm eff}}{\partial C} &=&
 C \left[\,\lambda_{BC}B^2 + \lambda_{CD} D^2\,\right] = 0,  
 \label{gap-eq-C} \\
 \frac{\partial V_{\rm eff}}{\partial D} &=&
 T B^2 + \lambda_{BD} D B^2 + \lambda_{CD} C^2 D =0 ,
 \label{gap-eq-D}
\end{eqnarray}
and 
\begin{equation}
  \frac{\partial V_{\rm eff}}{\partial \mu_e} = 0, \qquad
  \frac{\partial V_{\rm eff}}{\partial \mu_8} = 0 , \qquad
  \frac{\partial V_{\rm eff}}{\partial \Delta} = 0. 
\end{equation}

We can expand $\mu_e$, $\mu_8$, and $\Delta$ around
$B=C=D=0$,
\begin{eqnarray}
\label{bar1}
  \mu_e &=& \bar{\mu}_e + \xi_e, \\
  \mu_8 &=& \bar{\mu}_8 + \xi_8 , \\ 
  \Delta &=& \bar{\Delta} + \xi_\Delta,
\label{bar3}
\end{eqnarray}
where the bar-quantities are the 2SC solution, when $B=C=D=0$.

Let us assume that the origin (bifurcation point)
of the solution with nonzero
$B$, $C$, and $D$ corresponds to a second order phase transition 
(as will become clear in a moment, this assumption is self-consistent).
Then, taking an infinitesimally small $B$ near the critical point,
we easily find that
\begin{equation}
  \xi_e, \xi_8, \xi_\Delta \sim {\cal O}(B^2) .
  \label{xi}
\end{equation}
It implies that the difference of $V_\Delta$ in 
the new solution and that in the 2SC one is
\begin{equation}
   V_\Delta(\Delta^{\rm sol}, \mu_e^{\rm sol},\mu_8^{\rm sol})
 - V_\Delta(\bar{\Delta}, \bar{\mu}_e,\bar{\mu}_8) \sim {\cal O}(B^4) .
 \label{diff}
\end{equation}
This fact will be useful in our analysis below.

{}From Eqs. (\ref{gap-eq-B})--(\ref{gap-eq-D}),
we find that when the 2SC solution becomes unstable ($M_B^2 < 0$),
a new solution occurs, {\sl if} the parameters $\lambda_{BC}$ and
$\lambda_{CD}$ satisfy
\begin{equation}
  \lambda_{BC} > 0, \quad \lambda_{CD} < 0
\label{constraint}
\end{equation}
(in the next section, it will be shown that this constraint is
satisfied indeed). The new solution is:
\begin{equation}
  B_{\rm sol} = \frac{-M_B^2}{3|T|}
           \sqrt{\frac{-\lambda_{CD}}{\lambda_{BC}}}, \quad
  C_{\rm sol} =  \sqrt{\frac{-M_B^2}{3\lambda_{BC}}}, \quad
  D_{\rm sol} = \frac{-M_B^2}{3T},
  \label{app}
\end{equation}
where we neglected higher order terms of $M_B^2$. 
In Eq.~(\ref{app}) the conventions $B > 0$ and $C > 0$ are chosen.

Near the critical point $M_B^2 = 0$, 
the solution behaves as
\begin{equation}
  B_{\rm sol} \propto -M_B^2, \quad C_{\rm sol} \propto \sqrt{-M_B^2}, \quad
  D_{\rm sol} \propto -M_B^2 .
  \label{scaling}
\end{equation}
These scaling relations are quite remarkable.
While the scaling relation for $C$ is of engineering type, 
those for $B$ and $D$ are not (the origin of this is of
course in the presence of the 
dimensional coefficient $T$ in Eq. (\ref{app})).
Such a scaling behavior implies that the $B^4$ and $B^2 D^2$ terms  
in the effective potential are irrelevant
near the critical point $M_B^2 =0$. Omitting them,
we arrive at the reduced effective potential:
\begin{equation}
  \tilde{V}_{\rm eff} =
  V_\Delta + \frac{1}{2}M_B^2 B^2
  + T D B^2 + \frac{1}{2}\lambda_{BC} B^2 C^2
  + \frac{1}{2}\lambda_{CD} C^2 D^2 .
  \label{V_min}
\end{equation}

Notice that $(\tilde{V}_{\rm eff}-V_\Delta) \sim {\cal O}(B^3)$. This fact
and Eq. (\ref{diff}) imply that in the leading approximation
one can use the bar-quantities, defined in Eqs. (\ref{bar1})-(\ref{bar3}),
in calculating $V_\Delta$, $M_B^2$, $T$, $\lambda_{BC}$, and 
$\lambda_{CD}$ in the reduced potential. 
In other words,  
the effective potential can be decomposed into the ``constant'' 
2SC part $V_\Delta$, with frozen fermion parameters, and the dynamical 
gluonic part:
\begin{equation}
  \tilde{V}_{\rm eff} \to
  \tilde{V}_{\rm eff}(\bar{\Delta},\bar{\mu}_e,\bar{\mu}_8;B,C,D) = 
  V_\Delta(\bar{\Delta},\bar{\mu}_e,\bar{\mu}_8) + \frac{1}{2}M_B^2 B^2
  + T D B^2 + \frac{1}{2}\lambda_{BC} B^2 C^2
  + \frac{1}{2}\lambda_{CD} C^2 D^2 .
  \label{V_min1}
\end{equation}
Then Eq.~(\ref{app}) is the exact solution of the potential (\ref{V_min1}) 
and the energy density at the stationary point is found as
\begin{equation}
  \tilde{V}_{\rm eff}(\bar{\Delta},\bar{\mu}_e,\bar{\mu}_8;
             B_{\rm sol},C_{\rm sol},D_{\rm sol})
 =V_\Delta + \frac{1}{6}M_B^2 B^2_{\rm sol} = 
  V_\Delta - \frac{(-M_B^2)^3}{54 T^2}
  \left(\,-\frac{\lambda_{CD}}{\lambda_{BC}}\,\right) < V_\Delta .
\end{equation}
Therefore the gluonic vacuum is more stable than the 2SC one.

By using the Gauss's law constraint
\begin{equation}
  T B^2 + \lambda_{CD}C^2 D = 0,
\end{equation}
we find the true effective potential without the non-dynamical degree of
freedom $A^{(3)}_{0}$:
\begin{equation}
  \tilde{V}_{\rm eff}^{\rm Gauss} = V_\Delta + \frac{1}{2}M_B^2 B^2
  + \frac{1}{2}\lambda_{BC} B^2 C^2
  - \frac{T^2 B^4}{2\lambda_{CD}C^2} .
  \label{V_Gauss}
\end{equation}
It is easy to show that solution (\ref{app}) is
a minimum by analyzing the curvature of 
$\tilde{V}_{\rm eff}^{\rm Gauss}$. 

In the next section, we will calculate $M_B^2$, $T$, $\lambda_{BC}$,
and $\lambda_{CD}$. In particular, it will be shown 
that constraint (\ref{constraint}) is satisfied near the critical point.

\section{Dynamics in one-loop approximation}
\label{4}

In this section, we determine the GL effective potential (\ref{V_min1})
in one-loop approximation and determine the dispersion relations for
quarks in the gluonic phase. 
The 2SC $V_\Delta$ part of the potential is known~\cite{Huang:2003xd},
\begin{eqnarray}
  V_\Delta (\Delta,\mu_e,\mu_8) &=&
    \frac{\Delta^2}{4G_\Delta} - \frac{\mu_e^4}{12\pi^2}
  - \frac{\mu_{ub}^4}{12\pi^2} - \frac{\mu_{db}^4}{12\pi^2}
  - \frac{\tilde{\mu}^4}{3\pi^2}
    \nonumber \\ &&
  - \frac{\Delta^2}{\pi^2}\left[\,
     \tilde{\mu}^2-\frac{1}{4}\Delta^2\,\right]\ln \frac{4\Lambda^2}{\Delta^2}
  - \frac{\Delta^2}{\pi^2}\left[\,\Lambda^2 -
    2\tilde{\mu}^2+\frac{1}{8}\Delta^2\,\right] , \quad (\delta\mu < \Delta).
\end{eqnarray}
Here $\Lambda$ is the ultraviolet cutoff in the NJL model and
$\mu_{ub}$, $\mu_{db}$, and $\tilde{\mu}$ are given in 
Eqs.~(\ref{beta-eq}) and (\ref{mu}).
For clarity of the presentation, the  
bars in $\Delta$, $\mu_{e}$ and $\mu_{8}$ were omitted 
(${\cal O}(\tilde{\mu}^2/\Lambda^2)$ and 
${\cal O}(\Delta^2/\Lambda^2)$ and higher terms are neglected
in this expression).
Note that the color and electrical charge neutrality conditions 
in the 2SC solution yield~\cite{Huang:2003xd}
\begin{equation}
  \delta \mu = \frac{3}{10}\mu - \frac{1}{5}\mu_8, 
  \label{2SC1}
\end{equation}
and
\begin{equation}
  (\tilde{\mu}^2+\delta\mu^2)\mu_8 =
   -\tilde{\mu}\Delta^2\left(\,\ln\frac{2\Lambda}{\Delta}-1\,\right)
   +\tilde{\mu}(\delta\mu^2+\mu_8^2)-\frac{1}{3}\mu_8^3 ,
  \label{2SC2}
\end{equation}
which is consistent with the result of Ref.~\cite{GR}, 
$\mu_8 \sim {\cal O}(\Delta^2/\mu)$,
in the case of 
$\delta\mu=0$.
The size of $\Delta$ is essentially determined by tuning the NJL
coupling constant $G_\Delta$ and cutoff $\Lambda$.

After straightforward but tedious 
calculations of relevant one-loop diagrams from the fermion
determinant in Eq.~(\ref{V}), we find the following relations  
in the region $\delta\mu < \Delta$:
\begin{eqnarray}
  M_B^2 &=& 
    - \frac{\mu_8^2}{g^2}
    + \frac{\tilde{\mu}^2}{6\pi^2}
      \left(\,1-\frac{2\delta\mu^2}{\Delta^2}\,\right), 
  \label{M_B} \\[3mm]
 \lambda_{BC} &=& \frac{1}{80\pi^2}\frac{\tilde{\mu}^2}{\Delta^2}
             \left[\,-1+8\frac{\delta\mu^2}{\Delta^2}
             \left(\,1-\frac{\delta\mu^2}{\Delta^2}\,\right)\,\right],
 \label{bc}  \\[3mm]
 \lambda_{CD} &=&
  -\frac{1}{g^2}-\frac{1}{18\pi^2}\frac{\tilde{\mu}^2}{\Delta^2},
 \label{cd}  \\[3mm]
 T &=& \frac{\mu_8}{2g^2} + \frac{\mu_8}{24\pi^2}\frac{\tilde{\mu}^2}{\Delta^2} 
       \left(\,-1+8\frac{\delta\mu^4}{\Delta^4}\,\right)
      +\frac{\tilde{\mu}}{48\pi^2}
       \left(\,-1+4\frac{\delta\mu^2}{\Delta^2}
                 +8\frac{\delta\mu^4}{\Delta^4}\,\right) .
 \label{T}
\end{eqnarray}
Here the tree contribution of gluons
\begin{equation}
  V_g \equiv -{\cal L}_g = -\frac{1}{2}
  F_{0j}^{(\alpha)} F_{0j}^{(\alpha)} =
  -\frac{1}{2g^2}\mu_8^2 B^2 + \frac{1}{2g^2}\mu_8 DB^2
  - \frac{1}{8g^2}B^2 D^2 - \frac{1}{2g^2}C^2 D^2 
\end{equation}
was also taken into account. 

We see that the coefficient $\lambda_{CD}$ is definitely negative.
The parameter $M_B^2$, which is expressed through 
the Meissner mass (\ref{m_M4}), is negative when
\begin{equation}
  \delta \mu > \delta\mu_{\rm cr}, \qquad
  \delta\mu_{\rm cr} = \frac{\Delta}{\sqrt{2}}
   \sqrt{1-\frac{3\pi}{2\alpha_s}\frac{\mu_8^2}{\tilde{\mu}^2}},
  \qquad \alpha_s \equiv \frac{g^2}{4\pi} .
\label{cr}
\end{equation}
Relation (\ref{2SC1}) and Eq.~(\ref{mu}) yield
\begin{equation}
  \tilde{\mu}=\frac{9}{10}\mu+\frac{2}{5}\mu_8,
\label{3}
\end{equation}
and, at the critical point, we find from Eqs. 
(\ref{2SC1}), (\ref{2SC2}) and (\ref{cr}), (\ref{3}) 
that $\mu_{8}$ is approximately
\begin{equation}
  \mu_8 = \frac{3-\ln\frac{200\Lambda^2}{9\mu^2}}{12+\frac{4}{9}\left(\,
     \ln\frac{200\Lambda^2}{9\mu^2}-2\,\right)}\,\mu .
\end{equation}
For realistic values $\Lambda = (1.5-2.0) \mu$ and 
$\alpha_s=0.75-1.0$, we obtain numerically 
\begin{equation}
  \frac{3\pi}{2\alpha_s}\frac{\mu_8^2}{\tilde{\mu}^2} = 0.03\mbox{--}0.1\;.
\end{equation}
This implies that the tree gluon contribution decreases 
the value of $\delta\mu_{\rm cr}$ by 1.5\%--5\% in comparison to
its value in the (non-gauged) NJL model. The smallness of this
correction is in accordance 
with the dominance of hard-dense-loop diagrams.

Let us now turn to the coefficient $\lambda_{BC}$ (\ref{bc}). 
At the critical point 
$\delta\mu = \delta\mu_{\rm cr}$, it is:
\begin{equation}
 \lambda_{BC} = \frac{1}{80\pi^2}\frac{\tilde{\mu}^2}{\Delta^2}
\left(\,1-\frac{9\pi^2}{2\alpha_s^2}\frac{\mu_8^4}{\tilde{\mu}^4}\,\right).
\label{bc1}
\end{equation}
Because the $\mu_8^4/\tilde{\mu}^4$-term is negligibly small,
we conclude that 
the coefficient $\lambda_{BC}$ is positive near the critical point.
Thus, constraint (\ref{constraint}) is satisfied indeed. 

Utilizing Eqs.~(\ref{M_B})--(\ref{T}) in Eq.~(\ref{app}),
one can obtain the solutions for $B$, $C$, and $D$ in the 
near-critical region.
Indeed, neglecting higher order terms in $\mu_8/\mu$ in 
(\ref{M_B})--(\ref{T}),
we get the approximate relations
\begin{equation}
  M_B^2 \simeq \frac{\tilde{\mu}^2}{6\pi^2}
         \left(\,1-\frac{\delta\mu^2}{\delta\mu_{\rm cr}^2}\,\right), \quad
  \lambda_{BC} \simeq \frac{9}{160\pi^2}, \quad
  \lambda_{CD}\simeq -\frac{1}{4\pi\alpha_s}-\frac{1}{4\pi^2}, \quad
  T \simeq  \frac{\tilde{\mu}}{16\pi^2}
     +\frac{\mu_8}{16\pi^2}\left(\,3+\frac{2\pi}{\alpha_s}\,\right), 
\end{equation}
which lead us to the near-critical solution:
\begin{eqnarray}
B_{\rm sol} &=& 
        \frac{\delta\mu^2-\delta\mu_{\rm cr}^2}{\delta\mu_{\rm cr}^2}
        \frac{16\,\tilde{\mu}\,
              \sqrt{10\left(\,1+\frac{\pi}{\alpha_s}\,\right)}}
             {27\left[\,1+\frac{\mu_8}{\tilde{\mu}}
                 \left(\,3+\frac{2\pi}{\alpha_s}\,\right)\,\right]},
 \label{b} \\[3mm]
C_{\rm sol} &=& \frac{\sqrt{\delta\mu^2-\delta\mu_{\rm cr}^2}}
                    {\delta\mu_{\rm cr}}\,
              \frac{4\sqrt{5}\,\tilde{\mu}\,}{9},
 \label{c} \\[3mm]
D_{\rm sol} &=& 
        \frac{\delta\mu^2-\delta\mu_{\rm cr}^2}{\delta\mu_{\rm cr}^2}
        \frac{8\,\tilde{\mu}}
             {9\left[\,1+\frac{\mu_8}{\tilde{\mu}}
               \left(\,3+\frac{2\pi}{\alpha_s}\,\right)\,\right]}. 
 \label{d}
\end{eqnarray}

It is noticeable that this solution 
describes nonzero field strengths $F_{\mu\nu}^{(\alpha)}$
which correspond to the presence
of
{\it non-abelian}
constant chromoelectric-like condensates in the ground state:
\begin{eqnarray}
E_{3}^{(2)} &=& F_{03}^{(2)} = \frac{1}{g}\,C_{\rm sol}D_{\rm sol}\,,\\
E_{3}^{(7)} &=& F_{03}^{(7)} = \frac{1}{2g}\,B_{\rm sol}
                             \left(\,2\mu_{8} - D_{\rm sol}\,\right).
\end{eqnarray}
We emphasize that while
an abelian constant electric field in different
media always leads to an instability,
\footnote {In metallic and superconducting
media, such an instability is classical in its origin.
In semiconductors and insulators, this instability is
manifested in an creation of electron-hole
pairs through a quantum tunneling process.}
non-abelian constant
chromoelectric fields do not in many cases: For a thorough discussion
of the stability problem 
for constant $SU(2)$ non-abelian  fields in theories with 
zero baryon density, see Ref. \cite{BW}.  
On a technical side, this difference is
connected with that while a vector potential corresponding to
a constant abelian electric field depends on spatial and/or time
coordinates,
a constant non-abelian chromoelectric field is expressed through
constant vector potentials, as takes place in our case, and
therefore momentum and energy are good quantum numbers in the latter.

In order to illustrate the stability issue in the gluonic phase,
let us consider the dispersion relations for quarks there.
Because the vacuum expectation values (\ref{b})-(\ref{d}) are small
near the critical point and because red and green quarks are gapped
in the 2SC phase,
the dispersion relations for gapless blue up and down quarks
are of the most interest. 
{}From Eq. (\ref{S-inv}) we find
that up to the first order in $B^2$ they are
\begin{eqnarray}
  p^{0}_{ub} &=& |\vec p|-\mu_{ub}
  +\frac{B_{\rm sol}^2}{4}
   \frac{1}{2|\vec p|+\mu_8+\frac{\Delta^2}{2\tilde{\mu}-\mu_8}}
  -\frac{B_{\rm sol}^2}{4}\frac{(p^3)^2}{\vec p{\;}^2}\left(
    \frac{1}{2|\vec p|+\mu_8+\frac{\Delta^2}{2\tilde{\mu}-\mu_8}}
  +\frac{2(|\vec p|-\tilde{\mu})+\mu_8}
        {\Delta^2-\mu_8^2-2\mu_8(|\vec p|-\tilde{\mu})}\,\right),
 \label{bu} \\[3mm]
  p^{0}_{db} &=& |\vec p|-\mu_{db}
  +\frac{B_{\rm sol}^2}{4}
   \frac{1}{2|\vec p|+\mu_8+\frac{\Delta^2}{2\tilde{\mu}-\mu_8}}
  -\frac{B_{\rm sol}^2}{4}\frac{(p^3)^2}{\vec p{\;}^2}\left(
    \frac{1}{2|\vec p|+\mu_8+\frac{\Delta^2}{2\tilde{\mu}-\mu_8}}
  +\frac{2(|\vec p|-\tilde{\mu})+\mu_8}
        {\Delta^2-\mu_8^2-2\mu_8(|\vec p|-\tilde{\mu})}\,\right) .
 \label{bd}
\end{eqnarray}
The $B^2$-terms in Eqs.(\ref{bu}) and (\ref{bd}) 
lead to non-spherical Fermi surfaces determined by the
following equations: 
\begin{eqnarray}
 |\vec p| &=& \mu_{ub}
  -\frac{B_{\rm sol}^2\sin^2\theta}{4}
   \frac{1}{2\mu_{ub}+\mu_8+\frac{\Delta^2}{2\mu_{ub}+\mu_e+\mu_8}}
  -\frac{B_{\rm sol}^2\cos^2\theta}{4}
   \frac{\mu_e+\mu_8}
        {\Delta^2+\mu_8 \mu_e+\mu_8^2} , 
 \qquad \mbox{(blue up)}
 \label{fs-up} \\[3mm]
 |\vec p| &=& \mu_{db}
  -\frac{B_{\rm sol}^2\sin^2\theta}{4}
   \frac{1}{2\mu_{db}+\mu_8+\frac{\Delta^2}{2\mu_{db}-\mu_e+\mu_8}}
  +\frac{B_{\rm sol}^2\cos^2\theta}{4}
   \frac{\mu_e-\mu_8}
        {\Delta^2-\mu_8 \mu_e+\mu_8^2} , 
 \qquad \mbox{(blue down)}
 \label{fs-down}
\end{eqnarray}
where we neglected higher order terms of $B^2$ and 
defined the angle $\theta$,
\begin{equation}
  p^3 \equiv |\vec p|\cos\theta .
\end{equation}
The dispersion relations (\ref{bu}) and (\ref{bd}) clearly show that 
there is no instability in the quark sector in this problem.

As to bosonic degrees of freedom (gluons and composite bosons),
because it is very involved to derive their derivative terms
from the fermion loop in the gluonic phase,
this issue is beyond the scope of this letter. It is however noticeable
that there are no
instabilities for bosons in a phase with vector condensates in
the gauged $\sigma$-model with a chemical potential for hypercharge
\cite{GMS}. Although that model is much simpler than
the present one, its phase with vector condensates has many common
features with the gluonic phase and this fact is encouraging.

We emphasize that these constant color condensates in the gluonic
phase do {\it not} produce long range color {\it forces} acting on
quasiparticles.
This can be seen from the dispersion relations (68) and (69) for quarks
in this model. They show that momentum and energy are conserved numbers.
It would be of course impossible in the presence of long range forces.
The role of these condensates is actually more dramatic: They change
the structure of the ground state, making it anisotropic and
(electrically) superconducting. Only in this sense, one can speak about a
long range character of the condensates.

\section{More about dynamics in the gluonic phase}
\label{5}

In this section, we will describe some additional features of
the gluonic phase. In particular, we will point out that a condensation
of {\it exotic} vector mesons takes place in this dense medium.

The
gluonic vector condensates 
are mostly generated at energy scales between the
confinement scale in the 2SC state, which is
$\lesssim 10$ MeV, and the baryon chemical potential, 
which is about
300-500 MeV. It is the same region where the chromomagnetic instability in
the 2SC phase is created and where the hard dense loop approximation
is (at least qualitatively) reliable. 
At such scales, gluons are still
appropriate dynamical degrees of freedom and utilizing the Higgs approach
with color condensates in a particular gauge is appropriate and
consistent: It is a region of hard physics. Because the gluonic phase 
occurs as a result of 
a conventional second order phase transition, the
vector condensates are very small only in the immediate 
surroundings of the critical point 
$\delta\mu = \Delta /\sqrt{2}$. 
Outside that region, their values should be of the order of the
typical scale $\delta\mu \sim \Delta \sim 100$ MeV.   

These condensates
represent hard dynamics connected with
the appearance of
a new parameter, chemical potential for the electric charge
$\mu_e \sim 100$ MeV. As a result, 
the $SU(2)_c$ gauge symmetry becomes completely broken
and the strong coupling (confinement)
dynamics presented
in the 2SC solution at the scale of order 10 MeV is washed out.
In other words, a conventional Higgs mechanism is 
realized in the gluonic phase. In this respect, the gluonic phase 
is similar to the color-flavor locked (CFL) phase, where the constant
color condensates (although not vector ones) completely 
break the $SU(3)_c$ color gauge symmetry \cite{RW}. 

It is easy to check that the electric charge
$\tilde{\tilde{Q}}$ and the
baryon number $\tilde{\tilde{\cal B}}$ 
introduced in Section 3 are integer both for gluons and quarks.
Do they describe hadronic-like excitations? 
We believe that the answer is ``yes". 
The point is that in models like this one, with a Higgs field
in the fundamental representation of the gauge group, there is no
phase transition between Higgs and confinement phases
\cite{DRS}. These two phases
provide dual, and physically equivalent, descriptions of 
dynamics (the complementarity principle). In particular, they
provide two complementary descriptions of a spontaneous breakdown of 
global symmetries, such as the rotational SO(3) 
and the electromagnetic $U(1)$ in the present case.
Following Ref. \cite{DRS},
one can apply the dual gauge invariant
approach in this model and show that all the gluonic and quark fields 
can be replaced
by colorless composite fields. The flavor numbers of these fields
are described by the conventional 
electric and baryon charges $Q$ and $B$. They are integer and coincide
with those the operators $\tilde{\tilde{Q}}$ and 
${\tilde{\tilde{\cal B}}}$ yield for gluonic and quark fields. 

While these issues will be considered in more detail elsewhere, 
here we would like to
point out the following noticeable feature of these states: 
{\it some of them are exotic}. 
For
example, the electric and baryon charges  $\tilde{\tilde{Q}}$ and
${\tilde{\tilde{\cal B}}}$
of
$A^{(+)}_{\mu} = A^{(1)}_{\mu} + iA^{(2)}_{\mu}$ gluons
are equal to +1 and +2, respectively. 
Because $A^{(+)}_{3}$ gluons
are condensed in the gluonic phase, we conclude that 
in the dual gauge invariant description
this corresponds to a condensation of
{\it exotic} vector mesons. In this regard,
it is appropriate to mention that some authors speculated about
a possibility of a condensation of vector $\rho$ mesons in dense
baryon matter \cite{rho}. The dynamics in the gluonic phase 
yield a scenario even with a more unexpected condensation.

\section{Summary and discussions}

The gluonic phase whose existence was shown in this paper is
very different from all known phases in dense quark matter
discussed in the literature. Also, to the best of our knowledge, no
phase like that has been considered in condensed matter
physics.
One of its features is the presence of {\it non-abelian}
constant chromoelectric condensates in the 
ground state. 
They make the dynamics of the gluonic phase to be manifestly non-abelian.
\footnote {The role of constant chromomagnetic non-abelian fields in the
dynamics of color superconductivity was studied in Ref.
\cite{EKT}. It was shown that they could enhance the value of a
diquark condensate. Unlike the gluonic phase, where non-abelian
chromoelectric condensates are solutions of dynamical equations, the 
chromomagnetic
fields in \cite{EKT} are external.} 

Because most of the initial symmetries in this system 
(including the rotational
$SO(3)_{\rm rot}$ and the electromagnetic $U(1)$) are spontaneously
broken, the spectrum of excitations in the gluonic phase should
be very rich. In particular, there should be two gapless 
Nambu-Goldstone (NG) modes connected with the two broken generators
of the rotational group and one NG mode corresponding to the
broken electric charge (the latter mode will be absorbed into photon
field neglected in our model).
\footnote
{The spectrum of excitations in the gauged $\sigma$-model with hypercharge
chemical potential~\cite{GMS}, where vector condensates also occur,
strongly supports these expectations.} Another interesting feature
of the gluonic phase is that there are excitations corresponding
to exotic hadrons. We are planning to return to this issue elsewhere.

The solution (\ref{b})--(\ref{d}) described in this paper 
corresponds to a minimum of the effective potential. Whether or not
this minimum is global is an open question. We suspect that there
may exist a class of solutions with vector condensates.
In this regard, it is instructive to describe the 
Larkin--Ovchinnikov--Fulde--Ferrell
(LOFF) phase~\cite{LOFF}
from this point of view. It is easy to show that
the LOFF solution
with one plane wave along, say, third spatial 
coordinate can be gauge transformed
into a solution with a
usual (homogeneous) diquark condensate {\it and} a vector
condensate $\VEV{A^{(8)}_{3}}$. [Note that there is no such a
transformation in the case of the LOFF solution with two or
more plane waves \cite{BR}.] 
Unlike the present gluonic solution,
there are no non-abelian chromoelectric condensates in that case, and,
therefore, the LOFF dynamics is not genuinely non-abelian. Still,
because the LOFF solution has been recently used to cure a
chromomagnetic instability in the neutral 2SC phase \cite{GIR}, 
it would be
worth to study a possibility of extending the present solution
by including the VEV $\VEV{A^{(8)}_{3}}$. 
\footnote {A model for describing dynamics for larger 
$\delta \mu \simeq \Delta$, close to the edge of the gapless
phase \cite{Huang:2003xd}, has been recently considered 
in Ref. \cite{H}. It is
an abelian gauge model with a photon gauge field having
a constant VEV. By using an appropriate gauge
transformation, one can remove this VEV, with a cost of introducing 
a phase factor in the order parameter, as it takes place in the
LOFF state.}

Last but not least, it would be interesting to check the possibility
of the existence of a gluonic phase in three-flavor dense quark matter.
It is especially interesting because a chromomagnetic instability
has been recently revealed also in that case \cite{CFL}. 

\acknowledgments

We thank K. Rajagopal and I. Shovkovy for useful discussions. The work 
was
supported by the Natural Sciences and Engineering Research
Council of Canada.

\end{document}